\documentclass[twocolumn,prl,showpacs,superscriptaddress]{revtex4}
\usepackage{graphicx}
\usepackage{amsmath}
\usepackage{amsfonts}
\usepackage{amssymb}
\usepackage{enumerate}
\usepackage{longtable}
\usepackage{dcolumn}
\usepackage{bm}
\usepackage[colorlinks,bookmarks=false,citecolor=blue,linkcolor=blue,urlcolor=blue]{hyperref}
\usepackage{color}
\begin{document}
\title{Theory of magnetization plateaux in the Shastry-Sutherland model}

\author{J.~Dorier}
\affiliation{Institute of Theoretical Physics, \'{E}cole Polytechnique F\'{e}d\'{e}rale de Lausanne, CH 1015 Lausanne, Switzerland}
\author{K.~P.~Schmidt}
\email{schmidt@fkt.physik.uni-dortmund.de}
\affiliation{Lehrstuhl f\"ur theoretische Physik, Otto-Hahn-Stra\ss e 4,D-44221 Dortmund, Germany}
\author{F.~Mila}
\affiliation{Institute of Theoretical Physics, \'{E}cole Polytechnique F\'{e}d\'{e}rale de Lausanne, CH 1015 Lausanne, Switzerland}
\date{\rm\today}

\begin{abstract}
Using perturbative continuous unitary transformations, we determine the long-range interactions
between triplets in the Shastry-Sutherland model, and we show that an unexpected structure develops at low magnetization with plateaux progressively appearing at 2/9, 1/6, 1/9 and 2/15 upon increasing the inter-dimer coupling. A critical comparison
with previous approaches is included.
Implications for the compound SrCu$_2$(BO$_3$)$_2$ are also discussed: we reproduce the magnetization profile around localized triplets revealed by NMR, we predict the presence of a 1/6 plateau, 
and we suggest that residual interactions beyond the
Shastry-Sutherland model are responsible for the other plateaux below 1/3.
\end{abstract}


\maketitle

\begin{figure}
   \begin{center}
   \includegraphics[width=0.8\columnwidth]{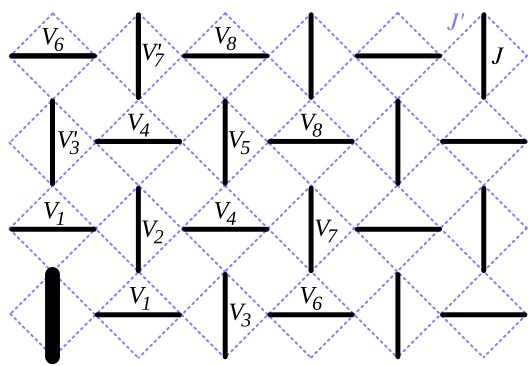}
   \end{center}
    \caption{Shastry-Sutherland lattice and definition of the 2-body interactions. $V_n$ is the coefficient of the 2-body interactions between the thick dimer and the dimer labeled  $V_n$.}
    \label{fig:lattice}
\end{figure}

Following the discovery of magnetization plateaux in the layered copper
oxide SrCu$_2$(BO$_3$)$_2$\cite{onizuka00,kodama02}, a lot of activity has been devoted to the 
properties in a magnetic field of the 2D spin-1/2 Heisenberg model 
known as the Shastry-Sutherland model\cite{shastry82}
and defined by the Hamiltonian:
\[H=J'\sum_{<i,j>}\bm S_{i}\cdot \bm S_{j}+J\sum_{\ll i,j\gg}\bm
S_{i}\cdot \bm S_{j}-B\sum_{i}S_i^z\quad,\] 
where the $\ll i,j\gg$ bonds build an array of orthogonal dimers while
the $<i,j>$ bonds are best seen as inter-dimer couplings (see Fig.~\ref{fig:lattice}).
For $J'/J$ smaller than a critical ratio of order 0.7, the ground state of the
model is exactly given by the product of dimer singlets, and the magnetization
process can be described in terms of polarized triplets $|t^1\rangle=|\uparrow\uparrow\rangle$
on the dimers interacting and moving
on an effective square lattice\cite{momoi00,miyahara03R}. These triplets can be described as hard-core bosons,
and the magnetization plateaux correspond to Mott insulating phases.

All theoretical approaches agree on the presence of magnetization
plateaux at 1/3 and 1/2\cite{momoi00,misguich01,miyahara00,miyahara03R,miyahara03}, in agreement with experiments\cite{onizuka00,sebastian07}.
However, the structure below 1/3 is rather controversial. On the experimental
side, the original pulsed field data have only detected two anomalies interpreted as
plateaux at 1/8 and 1/4\cite{onizuka00}, but the presence of additional phase transitions and of
a broken translational symmetry above the 1/8 plateau has been established by recent
torque and NMR measurements up to 31 T\cite{Takigawa07,levy08}. The possibility of additional plateaux has been pointed
out by Sebastian et al\cite{sebastian07}, who have interpreted their high-field torque measurements
as evidence for plateaux at $1/q$ with $2\le q\le 9$ and at $2/9$.
On the theoretical side, the situation is not settled either. The finite clusters 
available to exact diagonalizations prevent reliable predictions for high-commensurability
plateaux, and the accuracy of the Chern-Simons mean-field approach
initiated by Misguich {\it et al.}\cite{misguich01} and recently used by Sebastian {\it et al.}\cite{sebastian07} to explain additional
plateaux is hard to assess. The essential difficulty lies in the fact that, since plateaux 
come from repulsive interactions between triplets, an accurate determination of the low-density,
high-commensurability plateaux requires a precise knowledge of the long-range part of the
interaction, which could not be determined so far. 

In this Letter, we combine perturbative continuous unitary transformations (PCUTs)\cite{knetter00} with an analysis
of the effective hard-core boson model reformulated as a spin model to investigate the magnetization
process of the Shastry-Sutherland model. As we shall see, below 1/3, the results are well controlled 
up to quite large values of $J'/J$, providing
compelling
evidence in favour of a rich and unexpected plateau structure 
at small magnetization. 

The PCUT transforms the Shastry-Sutherland model into an effective model conserving the number of elementary triplets (triplons\cite{schmi03}). The relevant processes for the physics in a finite magnetic field have maximum total spin and total $S_z$. Other spin channels relevant for spectroscopic observables have been studied earlier\cite{knetter00_2}. The general form of the effective Hamiltonian obtained by the PCUT reads therefore:
\begin{equation*}
H_{\rm eff} =\sum_{n=2,4,6\cdots}\ \sum_{r_1,\cdots,r_n}C_{r_1,\cdots,r_n}b_{r_1}^\dag \cdots b_{r_{n/2}}^\dag b_{r_{n/2+1}}\cdots b_{r_n}
\end{equation*}
where the $r_i$'s are sites of the square lattice formed by the $J$-bonds, while
the hard-core boson operator $b_r^\dag$ creates a polarized triplet
$|t^1\rangle$ at site $r$. The coefficients $C_{r_1,r_2,\cdots,r_n}$ are 
obtained as a series in $J'/J$. 
%
We have kept all terms with up to 3 creation and annihilation operators and all  
4-body interactions ($n_{r_1}n_{r_2}n_{r_3}n_{r_4}$) that first appear 
at order $\leq 8$. For the 2-body interactions ($n_{r_1}n_{r_2}$), which dominate the physics at
low densities (see below), we keep more terms, namely those that first appear at order less
or equal to 10. The coefficients $C_{r_1,r_2,\cdots,r_n}$ are evaluated up to order 15 for the
2-body interactions and up to order 12 for the other terms, and they are then extrapolated 
using Pad\'{e} or DlogPad\'{e} extrapolants. The resulting Hamiltonian
contains more than 15000 processes. 

\begin{figure}
   \begin{center}
     \includegraphics[width=0.7\columnwidth]{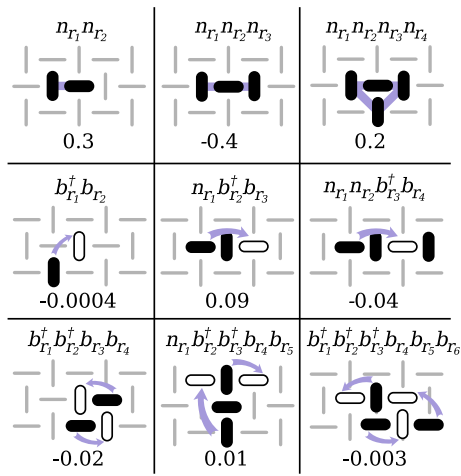}
   \end{center}
   \caption{Largest process of each type in the Hamiltonian $H_{\rm eff}$ and its amplitude $C_{r_1,\cdots,r_n}$ in units of $J$ at $J'/J=0.5$.}
    \label{fig:hamiltonien}
\end{figure}

Clearly it is not possible to describe all the terms of $H_{\rm eff}$. To give an idea of its structure, the
dominant terms of each type are sketched in Fig.~\ref{fig:hamiltonien}.  
In the range $0<J'/J<0.5$, the relative importance of the various terms 
(as deduced from their coefficients or from their contribution to the ground
state energy as estimated below) follows simple trends.
The pure density-density interaction terms have by far the largest coefficients. So one expects the magnetization curve
to consist mainly of plateaux separated by jumps, or by very narrow intermediate phases.
Secondly, the magnitude of all terms decreases when the distance between sites involved in the interaction 
increases. As a consequence, at low density, interaction terms involving more than two particles
give a negligible contribution to the energy, which is completely dominated by 2-body
interactions. Thirdly, the standard two-site hopping appears only at
order 6\cite{miyahara99} and is really small, a consequence of the frustration
of the model, and the kinetic part of the Hamiltonian is dominated
by the correlated hopping ($n_{r_1}b^\dag_{r_2}b_{r_3}$), which allows
a particle to hop only if there is another particle
nearby\cite{momoi00,knetter00_2}. Such a correlated hopping has been recently shown to strongly favour supersolid
phases\cite{schmidt08}, and supersolid rather than superfluid intermediate phases are 
possible.

\begin{figure}
   \begin{center}
     \includegraphics[width=0.9\columnwidth]{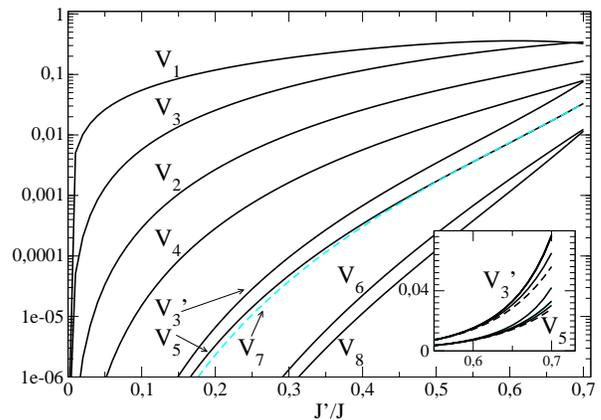}
   \end{center}
   \caption{Coefficients of the extrapolated 2-body interactions as a function of
     $J'/J$. Inset: Different extrapolants (solid lines) as well as the bare series (dashed lines) for $V'_3$
     and $V_5$.}
    \label{fig:hamiltonien_twobody_Jp}
\end{figure}

Given their importance at low density,
we show in Fig.~\ref{fig:hamiltonien_twobody_Jp} the evolution with
$J'/J$ of the
2-body interactions defined in Fig.~\ref{fig:lattice}. Although at low $J'/J$ 
the interactions smaller than $V_4$ may
be neglected, this is not true for larger $J'/J$ where the higher order
terms $V'_3$, $V_5$, and $V_7$ (appearing at order 6) become important and contribute to the
formation of low-density plateaux. For these terms, the bare series and
the Pad\'{e} extrapolations are basically indistinguishable below $J'/J=0.5$. Beyond that value and
up to $J'/J\simeq 0.63$,
various Pade extrapolations still give consistent results for these 
2-body interactions. 
Remarkably, we find that local interactions involving more than two particles become very strong for $J'/J\geq 0.6$ close to the phase transition and are very hard to extrapolate.   
We therefore restrict the discussion to $J'/J\le 0.5$ where 
the expansion is well controlled.
 
The effective Hamiltonian $H_{\rm eff}$ has 
both positive and negative off-diagonal terms, which prevents its investigation
by Quantum Monte Carlo (QMC). So, to determine its phase diagram, we map it onto a spin 1/2 model using the Matsubara-Matsuda representation\cite{matsubara56} of hard-core bosons $S^+=b$, $S^-=b^\dag$, $S^z=1/2-b^\dag b$, and we use a classical
approximation where the spins are treated as classical vectors of length
1/2.  
Given the complexity of the model, the energy is then minimized numerically on
finite size clusters with periodic boundary conditions. To allow for high commensurabilities
of various symmetries, all
non-equivalent clusters with up to 32 sites for all $J'/J$ (64 for $J'/J=0.5$)
have been tested and compared.
The possible phases are identified using symmetry arguments. A state is
superfluid if it breaks the U(1) gauge symmetry ($z$ axis spin
rotation), it is solid if it breaks the translationnal symmetry ($S^z$
is non-uniform), and it is a supersolid if it breaks both
symmetries. At the classical level, the relevant order parameters are
the in-plane component of the spins ($\sqrt{(S^x)^2+(S^y)^2}=|\langle
b\rangle |$) for the superfluid and the static structure factor for
the solid. This classical approximation has been shown to be remarkably good in cases
where its predictions could be compared with QMC\cite{schmidt08,schmidt08b}.
More precisely, this approach slightly overestimates the tendency to form plateaux but plateaux not present in this method are very unlikely to occur. 
 \begin{figure}
   \begin{center}
   \includegraphics[width=0.9\columnwidth]{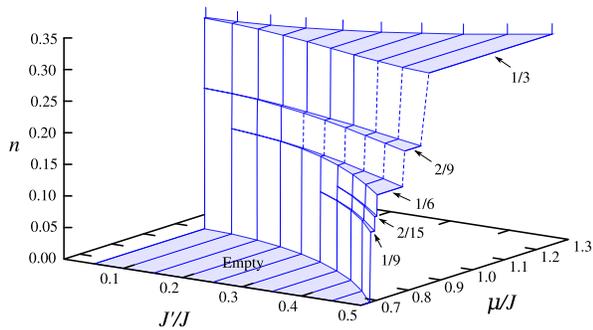}
   \end{center}
    \caption{Well converged plateaux as a function of $\mu$ and $J'/J$. The boson density $n$ is equal to
    the magnetization in units of the saturation value, and the chemical potential $\mu$ is equal to the magnetic field $B$.}
    \label{fig:phasediagram}
\end{figure}
\begin{figure}
   \begin{center}
   \includegraphics[width=0.8\columnwidth]{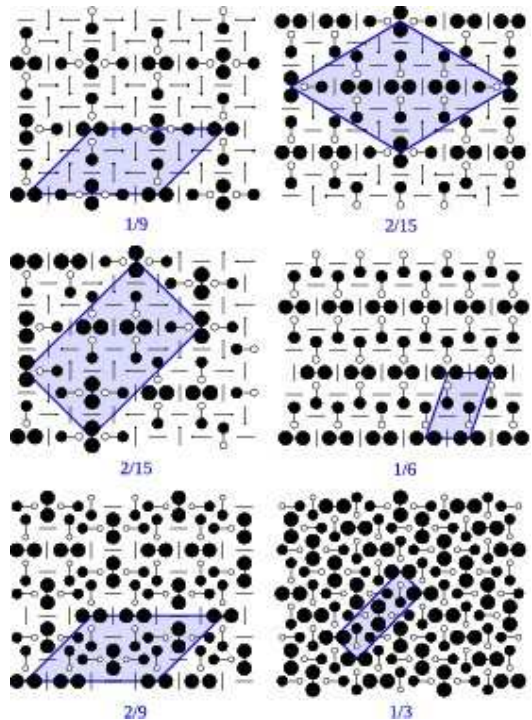}
   \end{center}
   \caption{Spin density ($S^z$) profile of the main plateaux
     at $J'/J=0.5$. Full (empty) circles corresponds to
     magnetization along (opposite to) the magnetic field. The radius of
     the circles is proportional to the magnetization amplitude. The
     blue line shows the unit cell compatible with the periodicity of
     the state. For the 2/15 plateau, two structures have the same energy within the error bars of the method.}
    \label{fig:plateaux}
\end{figure}

\begin{figure}
   \begin{center}
   \includegraphics[width=\columnwidth]{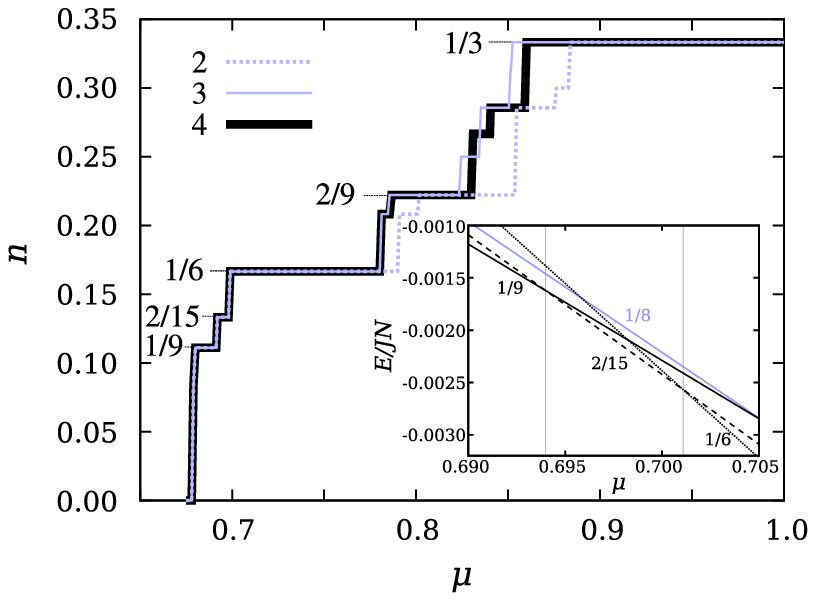}
   \end{center}
   \caption{Magnetization curve at $J'/J=0.5$: comparison of the results obtained by keeping terms
   with up to 2 creators (dotted blue curve), 3 creators (solid blue curve) and 4 creators (black curve). Well converged plateaux used in Fig.~\ref{fig:phasediagram} are indicated explicitly. Inset: Energy of the 1/9, rhomboid 1/8, 2/15, and 1/6 plateaux as a function of $\mu$ at $J'/J=0.5$. The error bars from the DlogPad\`{e} extrapolation are smaller than the line width.}
    \label{fig:convergence_NbCreators}
\end{figure}

The resulting phase diagram is shown in Fig.~\ref{fig:phasediagram}, and the structure
of the plateaux in Fig.~\ref{fig:plateaux}. 
The solid lines in the phase diagram denote results that are fully
converged with respect to the terms kept in the Hamiltonian (see Fig.~\ref{fig:convergence_NbCreators}). 
Well converged domains are then connected by dashed lines.
The phase diagram is dominated by a series of plateaux, at 1/3 and 1/2 (not shown)
already at very small $J'/J$, then by plateaux at 2/9, 1/6, 1/9 and 2/15. 
Whether these plateaux are separated by jumps or intermediate phases (with possibly additional plateaux) cannot be decided on the basis of the present calculation. 

The actual magnetization profile inside the various plateaux is depicted in Fig.~\ref{fig:plateaux}.
 The spin density profiles are obtained by transforming the relevant observable in the PCUT formalism\cite{knetter00}.
 In all cases, the building brick is a triplet and its two up-down neighboring dimers, in agreement
with the interpretation of Cu NMR in the first plateau of SrCu$_2$(BO$_3$)$_2$\cite{kodama02}.

Let us now compare the present results with previous works. Momoi and Totsuka
used pertubation theory to third order and logically found only plateaux at
1/3 and 1/2\cite{momoi00}. Miyahara and Ueda used a phenomenological form of the long-range
2-body interactions in a model without kinetic energy to successfully determine possible
structures inside various plateaux with emphasis on the plateaux at 1/8, 1/4 and 1/3 
reported in SrCu$_2$(BO$_3$)$_2$\cite{miyahara00}. However, their model exhibited several other plateaux,
and the approach was not set up to be predictive regarding the actual plateaux
stabilized in the model. Finally, a Chern-Simons theory where spins are mapped onto
fermions has been developed by Misguich {\it et al.}\cite{misguich01}, who only found plateaux at 1/4,
1/3 and 1/2, and by Sebastian {\it et al.}~\cite{sebastian07}, who allowed for non-uniform mean-field solutions
and found plateaux at $1/q$ with $q=2,...,9$ and at $2/9$. There are obvious similarities between the results of Sebastian et al and ours.
In particular, all our plateaux except the $2/15$ are present in their series, and this
exception could be due to the fact that
this plateau requires a unit cell not considered in Ref.~\onlinecite{sebastian07}. But some aspects of their results, for instance a well developed
plateau at 1/5, are definitely ruled out by our analysis and must be considered
as artefacts of the method. So further work is needed to understand the reliability and pitfalls of the
Chern-Simons approach to magnetization plateaux. 

Finally, let us discuss the implications for SrCu$_2$(BO$_3$)$_2$.
The most surprising aspect of our results is the absence
of plateaux at 1/8 and 1/4, which have been observed in SrCu$_2$(BO$_3$)$_2$.
This is especially so for the 1/8 plateau, which has not only been observed with
pulsed field, but has been extensively studied in steady fields with torque\cite{levy08}
and NMR\cite{kodama02}. To convince the reader that this plateau is indeed absent in the Shastry-Sutherland
model at $J'/J=0.5$, we have plotted in the inset of Fig.~\ref{fig:convergence_NbCreators} 
the energy of various low-density plateaux including the 1/8 plateau with rhomboid unit cell\cite{miyahara03}. 
Clearly, the 1/8 structure is never stabilized.

Taking for granted that the experimental identification of these 1/8 and 1/4 plateaux is correct (a point which
might actually deserve further investigation in view of the rather different absolute scales reported
in pulsed and steady magnetic field experiments), we can
think of two possible origins of this discrepancy. The first one is that the physics changes
dramatically when $J'/J$ approaches the critical value where the dimer product wave-function is no
longer the ground state but is replaced by a phase with plaquette order\cite{lauchli02}. This seems unlikely however
in view of the first order nature of this transition. 

The other possibility is that the plateau structure is influenced by residual couplings beyond the
simple SU(2) Shastry-Sutherland model. There are a priori three types of residual couplings, all of
the same order of magnitude ($J/100$): in plane Dyzaloshinskii-Moriya interactions (both
inter- and intra-dimer), further neighbor in plane exchange couplings, and inter-plane exchange 
couplings. In view of the very large parameter space generated by these couplings, a systematic 
study of their influence will be difficult. Recent ab-initio results might be helpful in restricting the
relevant portion of this parameter space though\cite{mazurenko08}. This is left for future investigation. Here it is interesting to notice that the 1/4 and rhomboid 1/8 plateau\cite{miyahara03} share with the 1/6 plateau the peculiarity to have all triplets parallel to each other, and to maximize the number of pairs of up-down dimers which are nearest neighbors on the same sublattice. Hence, if residual
interactions indeed favor the 1/8 (resp. 1/4) plateaux over the (1/9,2/15) pair (resp. 2/9), they are expected to further stabilize the 1/6 plateau, which thus appears as a very strong candidate for an additional plateau in SrCu$_2$(BO$_3$)$_2$.

To summarize, a systematic investigation of the effective Hamiltonian describing the triplets induced by
an external field in the Shastry-Sutherland model has lead to the prediction of an unexpected series of 
additional plateaux below 1/3 at 2/9, 1/6, 1/9 and 2/15 that appear progressively upon increasing the inter-dimer
coupling. High-order perturbation theory was essential in identifying the correct 
sequence of plateaux, a remark that opens interesting perspectives for other models. Regarding SrCu$_2$(BO$_3$)$_2$, our results suggest to have a closer look at the magnetization 
below 1/3, with two issues in mind: the absolute value of the magnetization at the plateaux currently 
assumed to be at 1/4 and 1/8, and the presence of a 1/6 plateau. 

We acknowledge very useful discussions with A.~Abendschein, C.~Berthier, S.~Capponi, M.~Horvatic, A.~Laeuchli, M.~Takigawa and G.~S.~Uhrig. KPS acknowledges ESF and EuroHorcs for funding through his EURYI. Numerical simulations were done on Greedy at EPFL ({\it greedy.epfl.ch}). This work has been supported by the SNF and by MaNEP.


\begin{thebibliography}{10}
\bibitem{onizuka00}
K. Onizuka {\it et al.}, J. Phys. Soc. Jpn. {\bf 69}, 1016 (2000).
\bibitem{kodama02}
K. Kodama {\it et al.}, Science {\bf 298}, 395 (2002).
\bibitem{shastry82}
B. S. Shastry and B. Sutherland, Physica B {\bf 108B}, 1069 (1981).
\bibitem{momoi00}
T. Momoi and K. Totsuka, Phys. Rev. B {\bf 62}, 15067 (2000).
\bibitem{miyahara03R}
S. Miyahara and K. Ueda, J. Phys.: Condens. Matter {\bf 15}, R327 (2003).
\bibitem{miyahara00}
S. Miyahara and K. Ueda, Phys. Rev. B {\bf 61}, 3417 (2000).
\bibitem{misguich01}
G. Misguich, T. Jolicoeur and S. M. Girvin, Phys. Rev. Lett. {\bf 87}, 097203 (2001).
\bibitem{miyahara03}
S. Miyahara, F. Becca and F. Mila, Phys. Rev. B {\bf 68}, 024401 (2003).
\bibitem{sebastian07}
S. E. Sebastian {\it et al.}, arXiv:0707.2075v1 (2007).
\bibitem{Takigawa07}
M. Takigawa {\it et al.}, arXiv:0710.5216v2 (2007).
\bibitem{levy08}
F. Levy {\it et al.}, Eur. Phys. Lett. {\bf 81}, 67004 (2008).
\bibitem{knetter00}
C. Knetter and G. S. Uhrig, Eur. Phys. J. B {\bf 13}, 209 (2000); C. Knetter, K.P. Schmidt, and G.S. Uhrig, J. Phys. A: Math. and Gen. {\bf 36}, 7889 (2003).
\bibitem{schmi03}
K.P. Schmidt and G.S. Uhrig, Phys. Rev. Lett. {\bf 90}, 227204 (2003).
\bibitem{knetter00_2}
C. Knetter, A. B\"uhler, E. M\"uller-Hartmann and G. S. Uhrig Phys. Rev. Lett. {\bf 85}, 3958 (2000); C. Knetter and G. S. Uhrig Phys. Rev. Lett. {\bf 92}, 027204 (2004).
\bibitem{miyahara99}
S. Miyahara and K. Ueda, Phys. Rev. Lett. {\bf 82}, 3701 (1999).
\bibitem{schmidt08}
K. P. Schmidt {\it et al.}, Phys. Rev. Lett. {\bf 100}, 090401 (2008).
\bibitem{matsubara56}
T. Matsubara and H. Matsuda, Prog. Theor. Phys. {\bf 16}, 569 (1956).
\bibitem{schmidt08b}
K.~P.~Schmidt, J.~Dorier, and A.~L\"auchli, arXiv:0805.1408.
\bibitem{lauchli02}
A. L\"auchli, S. Wessel, and M. Sigrist, Phys. Rev. B {\bf 66}, 014401 (2002).   
\bibitem{mazurenko08}
V.~V.~Mazurenko {\it et al.}, arXiv:0804.4771.
\end{thebibliography}
\end{document}